\documentclass[12pt,a4paper]{iopart} 
\usepackage{iopams}
\usepackage{setstack,cite,doublespace}
\usepackage{amssymb,graphicx}

\newfont{\logo}{logo10}

\newcommand{\bea}{\begin{eqnarray}}
\newcommand{\eea}{\end{eqnarray}}

\begin{document}
\bibliographystyle{revtex}
\title[]{Motion of Space Curves in Three-dimensional Minkowski
Space $R_1^{3}$, SO(2,1) Spin Equation and Defocusing Nonlinear
Schr\"odinger Equation} 

\author{GOPAL MUNIRAJA}
\address{Department of Mathematics, Bishop Cotton Women's Christian College, Bangalore-560027, India.}
%\ead{muni_bishop@yahoo.com}
\author{M LAKSHMANAN} 
\address{Center for Nonlinear Dynamics, Bharathidasan University,Tiruchirapalli 620 024, India }
      
\begin{abstract}
 We consider the dynamics of  moving curves in
 three-dimensional Minkowski space $R_1^{3}$ and deduce the
 evolution equations for the curvature and torsion of the curve.
 Next by mapping a continuous SO(2,1) Heisenberg spin chain on the
 space curve in $R_1^{3}$, we show that the defocusing nonlinear
 Schr\"odinger equation(NLSE)  can be identified with the spin chain,
 thereby giving a geometrical interpretation of it. The associated
 linear eigenvalue problem is also obtained in a geometrical
 way.
\end{abstract}

%\ams{1}
% \keywords %
%\maketitle
\section{INTRODUCTION}
Modelling of physical systems by curves, surfaces and other
differential geometric objects is highly rewarding, see for
example the pioneering work of Hasimoto [1] on vortex filaments
and the one-dimensional continuum Heisenberg ferromagnetic spin
equation by Lakshmanan, Ruijgork and Thompson [2,3]. In both cases
the systems were shown to be equivalent to the integrable
nonlinear  Schr\"odinger equation (NLSE) of the focusing type [1-5]. In
recent times  the relation between differential geometry and
certain dynamical systems described by nonlinear evolution
equations in (1+1) and (2+1) dimensions, especially the integrable
systems , has come into sharp focus [6-13].

Integrable nonlinear evolution equations occur in many branches of
physics and applied mathematics. Such equations possess a number
of interesting  properties such as soliton solutions, infinite
number of conservation laws, infinite number of symmetries, B\"acklund
and Darboux transformations, bi-Hamiltonian structures and so on,
see [7, 8].

Now it is well known that a class of important soliton equations
can be interpreted in terms of moving space curves in $R^{3}$ and
the linear eigen value problems of the soliton equations can be
obtained from the defining Serret-Frenet equations of space curves
 [6, 9-13].  Extension to (2+1) dimensions is also possible. A brief
survey of the developments up to more recent times can be seen in
[13].

Our focus in this Letter is on the defocusing NLSE, in which the
sign of the nonlinear term is negative, which is encountered in
many physical problems. It was shown to be integrable by the
inverse scattering transform method[14] and admits dark soliton
solutions. However, there does not seem to be available a simple
differential geometric model and its equivalent spin system for the defocusing case such as
available for the focusing NLSE as demonstrated in [1,2].  

In this Letter we wish to address this problem. We have found the
rich curve theory available in three-dimensional Minkowski space
is more suited to this problem. Nakayama [16] has used the
geometry of curves on a three-dimensional ellipsoid in a
four-dimensional Minkowski space to obtain a  model for the
defocusing NLSE.  For other related works see also [17,18].  But we feel our approach is simpler and more
direct and can be extended profitably to other nonlinear evolution
equations also. 

In the next section we give the basic equations of curves in
R${_1^3}$and fix the notation closely following [19] and [20].
Subsequent sections establish the curve model of the solution of
the defocusing NLSE with the tools and techniques similar to the
ones found in [3] and [20-23] and its connection to the SO (2,1) continuous Heisenberg spin chain.

We hope to extend similar treatment of integrable and
nonintegrable systems to surfaces in R${_1^3}$  and higher
dimensional Minkowski geometries in our subsequent work.
\section{\textbf{Motion of curves in  the Minkowski Space R${_1^3}$}}
The nature of the metric in a Minkowski space induces a rich
geometry of curves and surfaces. For instance the familiar
Serret-Frenet equations in the Euclidean space R${^3}$ give way to
four such systems in R${_1^3}$. In this section we give the basic
curve geometry apparatus in R${_1^3}$, see [19],  and we closely
follow [20] in writing down the curve evolution equations.  
The metric on the Minkowski space R${_1^3}$ is given by $ds^{2}$ =
-dx${_1^2}$ + dx${_2^2}$ + dx${_3^2}$.  We note here that the scalar and vector products of two vectors $a= a_{1}i+a_{2}j+a_{3}k$, $b= b_{1}i+b_{2}j+b_{3}k$ in R${_1^3}$,
where i, j, k are unit vectors along the x, y, z axes
respectively, are given as follows:

Scalar Product: $a.b= g(a, b)= -a_{1}b_{1}+a_{2}b_{2}+a_{3}b_{3}$.

Vector Product: $a\wedge b = \left|\begin{array}{ccc}
   -i &j & k\\
  a_{1} & a_{2} & a_{3} \\
  b_{1} & b_{2}& b_{3}\\
\end{array} \right|$,\\
( \emph{Note}: In Minkowski space a vector $a$ is
defined to be a unit vector if $g(a,a)= \pm 1$. A vector $a$ is
said to be \emph{space like} if
$g(a,a)>0$, \emph{time like} if $g(a,a)<0$ and \emph{light like} or a \emph{null vector }if $g(a,a)=0$  ).

Now, let($e_{1}$, $e_{2}$, $e_{3}$) be the Serret-Frenet frame of
a unit speed (non-null) curve $\alpha(x)$ in R${_1^3}$. Here
$e_{1}$ is the unit tangent vector field , $e_{2}$ is the normal
and $e_{3}$ is the binormal to $\alpha(x)$.  Let g($e_{1}$, $e_{1}$) = $\epsilon{_0}$ = $\pm 1$,
g($e_{2}$,$e_{2}$) = $\epsilon{_1}$ = $\pm 1$.
 Then g($e_{3}$,$e_{3}$) = -$\epsilon{_0}\epsilon{_1}$.  Then the Serret-Frenet  equations are given by [19]
\bea
e_{1}{_x}=\epsilon{_1}\kappa(x)e_{2},\nonumber\\
e_{2}{_x}=-\epsilon{_0}\kappa(x)e_{1}-\epsilon{_0}\epsilon{_1}\tau(x)e_{3},\nonumber\\
e_{3}{_x}=-\epsilon{_1}\tau(x)e_{2}.
\label{serret}
\eea
Here $\tau$ and $\kappa$ denote the torsion and curvature
respectively of the given space curve $\alpha$. Using the vector product relations
\bea
e_{1} \wedge e_{2}=e_{3},\nonumber\\
 e_{2} \wedge e_{3}=-\epsilon{_1}e_{1},\nonumber\\
e_{3} \wedge e_{1}=-\epsilon{_0}e_{2},
\eea
the Serret-Frenet equations (\ref{serret}) can be compactly written as
\begin{equation}
e_{ix}=D \wedge e_{i}, i = 1, 2, 3,
\end{equation}
where \emph{D} is the Darboux vector defined as
\begin{equation}
D=-\epsilon{_0}\epsilon{_1}\tau
e_{1}-\epsilon{_0}\epsilon{_1}\kappa e_{3}.
\end{equation}
Now, let us consider the time evolution of the curve
$\alpha(x,t)$. We define an angular momentum like vector $\Omega =
\Sigma\omega{_i}e_{i}$, i = 1,2,3, which gives the time evolution
of the Serret-Frenet system as
\begin{equation}
e_{i}{_t}=\Omega\wedge e_{i},    i=1,2,3.
\end{equation}
From (2) and (5) we obtain
\bea
e_{1t}=-\epsilon_{0}\omega_{3} e_{2}-\omega_{2} e_{3}, \nonumber\\
e_{2t}=\epsilon_{1}\omega_{3} e_{1}+\omega_{1} e_{3},\nonumber\\
e_{3t}=-\epsilon_{1}\omega_{2} e_{1} +
\epsilon_{0}\omega_{1}e_{2}.
\eea
In order that the above two definitions are compatible we require
that
\begin{equation}
({e}{_i}){_x}{_t}=({e}{_i}){_t}{_x}, i=1,2,3.
\end{equation}
From (1), (6 ) and (7) we obtain
\bea
\kappa_{t}=\tau\omega_{2}-\epsilon_{0}\epsilon_{1}\omega_{3x}, \nonumber\\
\tau_{t}=\epsilon_{1}\kappa\omega_{2}-\epsilon_{0}\epsilon_{1}\omega_{1x}, \nonumber\\
\omega_{2x}=\epsilon_{1}\tau\omega_{3}-\epsilon_{1}\kappa\omega_{1}.
\eea
The above equations constitute the evolution of the curvature and
torsion associated with an arbitrary curve moving in R${_1^3}$.
\section{\textbf{SO(2,1) Heisenberg Spin Equation and  Mapping to a Space Curve in R${_1^3}$}}
Consider now the SO(2,1) Heisenberg spin equation given by
\begin{equation}
S_{t}= S\times S_{xx},
\end{equation}
where $S$ is a unit vector in $R{_1^3}$, that is
$-S_{1}^{2}+S_{2}^{2}+S_{3}^{2}=\pm 1$ . We identify $S$ with the
unit tangent vector $e_{1}$ of $\alpha(x)$. Then we obtain from
the spin equation
\bea
e_{1t}&&=e_{1}\times e_{1xx}=e_{1}\times  e_{(1x)x}\nonumber\\
&&=e_{1}\times (\epsilon_{1}\kappa e_{2})_{x}=\epsilon_{1}e_{1}\times (\kappa_{x}e_{2}+ \kappa e_{2x})\nonumber
\eea
Hence from (1) and (2) and noting $\epsilon_{0}^{2}=\epsilon_{1}^{2}=1$ we have
\begin{equation}
e_{1t}=\epsilon_{1}\kappa_{x}e_{3} -\kappa \tau e_{2}.
\end{equation}
Next we have $e_{2}=\epsilon_{1}\frac{e_{1x}}{\kappa}$ from (1).  Hence
$e_{2t}= \epsilon_{1}\frac{e_{1xt}}{\kappa}-\frac{e_{1x}}{\kappa^{2}}\kappa_{t}.$  Using (1) and (10) we obtain
\begin{equation}
e_{2t}=[\epsilon_{1}\epsilon_{0}\kappa^{2}\tau
e_{1}-\epsilon_{1}(2\kappa_{x}\tau+\kappa\tau_{x}+\epsilon_{1}\kappa_{t})e_{2}+(\kappa_{xx}+\epsilon_{0}\kappa\tau^{2})e_{3}]/\kappa.
\end{equation}
Similarly we can deduce that
\begin{equation}
e_{3t}=
\kappa_{x}e_{1}+[(\epsilon_{0}\kappa_{xx}+\kappa\tau^{2})e_{2}-\epsilon_{1}(2\kappa_{x}\tau+\kappa\tau_{x}+\epsilon_{1}\kappa_{t})e_{3}]/\kappa.
\end{equation}
Comparing the above with (6) we immediately obtain the evolution equation for the curvature as 
\begin{equation}
\kappa_{t}=-\epsilon_{1}(2\kappa_{x}\tau+\kappa\tau_{x}).
\end{equation}
The compatiblity condition(7) applied to $e_{3}$ yields the
evolution equation for torsion as
\begin{equation}
\tau_{t}=-\epsilon_{1}^{2}\kappa\kappa_{x}-\epsilon_{0}\epsilon_{1}(\frac{\kappa_{xx}}{\kappa}+\epsilon_{0}\tau^{2})_{x}.
\end{equation}
The above equations define the evolution of curvature and torsion of the curve associated with an SO (2,1) continuum Heisenberg spin system in $R_1^3$.
\section{\textbf{Mapping onto the  Defocusing NLSE}}
Let us first consider the case where $\epsilon_{0}=-1$ and $\epsilon_{1}=1$.
Then equations (11) and (12) reduce to
\begin{equation}
\kappa_{t}=-2\kappa_{x}\tau-\kappa\tau_{x}\label{kt}
\end{equation}
 and
\begin{equation}
\tau_{t}=-\kappa\kappa_{x}+(\frac{\kappa_{xx}}{\kappa}-\tau^{2})_{x}.\label{tt}
\end{equation}
We now make the  complex transformation
\begin{equation}
u=\frac{\kappa}{2}e^{i\int^{x}_{-\infty}\tau dx}.\label{u}
\end{equation}
Then using (\ref{u}) equations (\ref{kt}) and (\ref{tt}) are transformed into
\begin{equation}
iu_{t}+u_{xx}-2|u|^2 u=0,
\end{equation}
which is nothing but the defocusing nonlinear Schr\"odinger
equation. Now we assume the energy and current densities of the spin system
to be related to the curvature and torsion respectively as
\bea
\epsilon(x,t)=\frac{1}{2}\frac{\partial S}{\partial
x}.\frac{\partial S}{\partial x }= \frac{1}{2}\kappa^{2},\\
I(x,t)=S.S_{x}\wedge S_{xx}
\eea
so that the continuity given by $\epsilon_{t}-I_{x}=0$ is
satisfied. This continuity equation can be easily shown to be
compatible with (\ref{kt}).  Finally we also observe that the case of $\epsilon_{0}=1$ and $\epsilon_{1}=-1$ yields the solution of focusing NLSE under the
condition that the curve has a constant torsion. The other two
cases of $\epsilon_{0}=-1$ , $\epsilon_{1}=-1$ and
$\epsilon_{0}=1$ , $\epsilon_{1}=1$ do not reduce either to the
defocusing or to the focusing NLSE for the transformation given by
(\ref{u}).
\section{\textbf{Reduction to AKNS Eigenvalue Problem}}
Corresponding to the Serret-Frenet frame  given by (1) (for
$\epsilon_{0}=-1$ and $\epsilon_{1}=1$) we define a new scalar
variable $z_{l}= \frac{e_{2l}+ie_{3l}}{1-ie_{1l}}$, l=1,2,3,
following [20, 21], from which we obtain
\begin{equation}
 z_{lx}= -i\tau z_{l}+\frac{i\kappa}{2}(1+z_{l}^{2}).\label{zlx}
\end{equation}
Now differentiating $z_{l}$ with respect to  $t$ and using (5), and 
after some detailed calculations, we arrive at
\begin{equation}
z_{lt}=-i\omega_{1}z_{l}+\frac{\omega_{2}+i\omega_{3}}{2}-\frac{(\omega_{2}-i\omega_{3})z_{l}^{2}}{2}.\label{zlt}
\end{equation}
Equations (\ref{zlx}) and (\ref{zlt}) are nothing but the Riccati equations.
Again the compatibility of (\ref{zlx}) and (\ref{zlt}) , that is $
(z_{l})_{xt}=(z_{l})_{tx}$, leads to the correct equations for
$\kappa(x,t)$ and $\tau(x,t)$ as in (\ref{kt}) and (\ref{tt}). Defining  $ z_{l}=\frac{v_{2}}{v_{1}}$,   equation (\ref{zlx}) can be
written as
\bea
v_{1x}= \frac{i\tau}{2}v_{1}-\frac{i\kappa}{2}v_{2},\nonumber\\
v_{2x}=\frac{i\kappa}{2}v_{1}-\frac{i\tau}{2}v_{2}.
\label{vx}
\eea
Similarly, from Eq.(\ref{zlt}) we  obtain
\bea
v_{1t}= \frac{i\omega_{1}}{2}v_{1}-\frac{i}{2}(\omega_{3}+i\omega_{2})v_{2},\nonumber\\
v_{2t}=-\frac{i\omega_{1}}{2}v_{2}+\frac{i}{2}(\omega_{3}-i\omega_{2})v_{1}.
\label{vt}
\eea
Using the compatibility conditions $(v_{ix})_{t}= (v_{it})_{x},
i=1, 2,$ from (\ref{vx}), (\ref{vt}) above we once again get back   easily
the original equations for $\kappa(x,t)$ and $\tau(x,t)$ .  Now introducing a suitable Galilean transformation and a gauge transformation in to (\ref{vx}), (\ref{vt}) we obtain the linear eigenvalue problem
\bea
\psi_{1x}=u \psi_{2}-i \lambda \psi_1, \nonumber\\
\psi_{2x}=u^* \psi_1+i \lambda \psi_2,
\eea
and the time evolution of the eigenfunction as
\bea
\psi_{1t}= A \psi_1+B \psi_2,\nonumber\\
\psi_{2t}=C \psi_1+D \psi_2,
\eea
where
\bea
 A=-2i\lambda^2-i u u^*, \nonumber\\
B=2u \lambda+i u_x, \nonumber\\
C=2u^* \lambda-i u^*_x.
\eea
and $u$ is as defined in Eq. (17).
\section{Conclusions}
In this paper, we have shown how the dynamics of moving curves in three dimensional Minkowski space $R_1^3$ can be related to the dynamics of SO(2,1) spin equations and soliton equations of defocusing NLS type.  It is possible that more number of such connections with soliton equations not related to moving curves in Euclidean space $R^3$ may be related to present type of formulation. The analysis can also be extended higher dimensional spaces.  Analysis along these lines is in progress.
 \section{\textbf{Acknowledgments}}
The first author (G. M.) was supported by a Summer Research Fellowship
of the Indian Academy of Sciences. The work of the second author (M. L.) was supported by a Department of Science and Technology (DST) Ramanna Fellowship and forms part of a DST-IRHPA research project.\\
\section*{References} 

\begin{enumerate}
\item {H. Hasimoto, A soliton on a vortex filament, J.Fluid.Mech.
\textbf{51}, (1972) 477-485}.

\item M. Lakshmanan,  Th. W. Ruijgork and C. J. Thompson, On the
dynamics of a continuum spin system, Physica A \textbf{84},
(1976) 577-590.

\item  M. Lakshmanan, Continuum spin system as an exactly solvable
dynamical system, Phys.Lett.A \textbf{61}, (1977) 53-54. 

\item  L. A. Takhtajan, Integration of the continuous Heisenberg spin chain through inverse scattering method, Phys. Lett. A \textbf{64}, (1977) 235-238.

\item V. E. Zakharov and  L. A. Takhtajan, Equivalence of the nonlinear Schr\"odinger equation and the equation of a Heisenberg ferromagnet, Theor. Math. Phys. \textbf{38}, (1979) 17-22.

\item G. L. Lamb Jr, Solitons on moving space curves,
J.Math.Phys.\textbf{18}, (1977) 1654-1661.

\item M. J. Ablowitz and P. A. Clarkson, Solitons, Nonlinear Evolution
Equations and Inverse Scattering, Cambridge University Press,
Cambridge (1992).

\item M. Lakshmanan, (Ed), Solitons, Springer-Verlag, Berlin (1988).

\item M. Lakshmanan,  On the geometric interpretation of solitons.
Phys.Lett.A  \textbf{64}, (1978) 353-356.

\item M. Lakshmanan, Rigid body motions, space curves, prolongation
structures, fiber bundles and solitons. J.Math.Phys. \textbf{20},
(1979) 1667-1672.

\item J. Cieslinski, P. K. Gragert and A. Sym, Exact solution to
localized-induction-approximation equation modeling smoke ring
motion. Phys.Rev.Lett. \textbf{57},(1986) 1507-1510.

\item J. S. Langer and R. Perline, Poisson geometry of the filament
equation. J.Nonlinear Sci. \textbf{1}, (1991) 71-93.

\item S. Murugesh and  M. Lakshmanan, Nonlinear dynamics of curves and
surfaces: Applications to physical systems, Int. J. Bifur. Chaos,
\textbf{15 },(2005) 51-63.
 
\item  V. E. Zhakarov and A. B. Shabat, Interaction between solitons in
stable medium, Sov. Phys. JETP. \textbf{37},
(1973) 823-828.

\item K. Nakayama,  H. Segur and M. Wadati, Integrability and the motion of curves, Phys.Rev.Lett. \textbf{69}, (1992) 2603-2606.

\item  K. Nakayama, Motion of curves in hyperboloid in the Minkowski
space, J. Phy. Soc. Japan. \textbf{67}, (1998) 3031-3037.

\item Q. Ding,  The gauge equivalence of the NLS and Schr\"odinger flow of maps in (2+1) dimensions, J. Phys. A \textbf{32}, (1999) 5087-5096.

\item N. G\"urb\"uz, The differential formula of Hasimoto transformation in Minskowski 3-space, Int. J. of Mathematics and Mathematical Sciences \textbf{16}, (2005) 2609-2616.

\item W. K\"uhnel, Differential Geometry: curves-surfaces-manifolds ,American Mathematical Society, Providence, RI (2006).

\item S. Vijayalakshmi, Nonlinear dynamics of higher dimensional spin systems,  Ph.D Thesis, Centre for Nonlinear Dynamics, Bharathidasan university, Tiruchirappalli, Tamil Nadu, India (2004).

\item M. Lakshmanan and S. Vijayalakshmi, Motion of curves and surfaces and nonlinear evolution equations in (2+1) dimensions in nonlinear phenomena in biological and physical sciences, (Eds.) S. K. Malik, M. K. Chandrasekar and N. Pradhan, 989-1013, Indian National Science Academy, New Delhi, (2000).

\item M. Lakshmanan,  R. Myrzakulov, S. Vijayalakshmi, and A. K. Danlybaeva,
 Motion of curves and surfaces and nonlinear evolution
equations in (2+1) dimensions, J. Math. Phys. \textbf{39},
(1998) 3765-3771.

\item R. Myrzakulov,  S. Vijayalakshmi, R. N. Syzdykova,  and M. Lakshmanan,
 On the simplest (2+1) dimensional integrable spin systems and
their equivalent nonlinear Schr\"odinger equations, J. Math.
Phys. \textbf{39}, (1998)2122-2140.
\end{enumerate}

\end{document}